\documentclass[aps,showpacs, pra,amssymb, amsmath, twocolumn]{revtex4}
\usepackage{amsmath,latexsym,amssymb}
\usepackage{graphicx}
\usepackage{amsmath}
\usepackage{latexsym}
\usepackage{amssymb}
\usepackage{bm}

\begin{document}

\title{Ultrafast Rabi oscillation of a Gaussian atom ensemble}
\author{Han-gyeol Lee, Hyosub Kim, and Jaewook Ahn}
\email{jwahn@kaist.ac.kr}
\address{Department of Physics, KAIST, Daejeon 305-701, Korea}
\date{\today}

\begin{abstract}
We investigate Rabi oscillation of an atom ensemble in Gaussian spatial distribution. By using the ultrafast laser interaction with the cold atomic rubidium vapor spatially confined in a magneto-optical trap, the oscillatory behavior of the atom excitation is probed as a function of the laser pulse power. Theoretical model calculation predicts that the oscillation peaks of the ensemble-atom Rabi flopping fall on the simple Rabi oscillation curve of a single atom and the experimental result shows good agreement with the prediction. We also test the  the three-pulse composite interaction $R_x(\pi/2)R_y(\pi)R_x(\pi/2)$ to develop a robust method to achieve a higher fidelity population inversion of the atom ensemble.
\end{abstract}

\pacs{ 32.80.Qk, 32.80.Wr, 42.65.Re}

\maketitle

Rabi oscillation is a fundamental concept in physics with a 
significant pedigree first discovered in the context of nuclear magnetic resonance (NMR)~\cite{RabiPR, RabiPR2, AbragamBook} and later extended to atomic physics and quantum optics~\cite{AllenBook, ScullyBook}. In the presence of an oscillatory driving field $E(t)=A(t) \cos(\omega t)$, a two-state quantum system undergoes a cyclic change of Bloch vector $\rho$  manifested by the precession 
\begin{equation}
{d\rho}/{dt} = \Omega \times \rho
\end{equation}
about an effective torque $\Omega=(-\mu A(t)/2\hbar, 0, \delta)$, where $\mu$ is the transition dipole moment between the two energy states, 
$A(t)$ is the field envelope, and  $\delta$  is the frequency detuning under the slowly-varying envelope approximation~\cite{AllenBook}.
This generic feature of Rabi oscillation is universally found in a vast variety of material systems ranging from simple atoms and molecules~\cite{JaynesCummings, Atom1, Atom2, Molecule1, Molecule2} to bulk semiconductors~\cite{Semiconductors}, quantum wells and dots~\cite{QW, QW2, QD, QD2}, graphene~\cite{Graphene}, surface plasmons~\cite{Plasmon}, superconducting interference devices~\cite{Superconductor, JosephsonJuntion}, diamond nitrogen-vacancy centers~\cite{Diamond}, and Bose-Einstein condensates~\cite{BEC}, etc. 

When a two-state atom  interacts with a resonant ($\delta=0$) laser pulse, the dynamics of the excited state probability, which we may refer to as single-atom Rabi oscillation (SARO), is represented by
\begin{equation}
P(\Theta_o)=\sin^2\frac{\Theta_o}{2}, \label{SARO}
\end{equation}
where $\Theta_o$ is the pulse area defined by $\Theta_o = \int \mu A(t) dt/\hbar$. Since the pulse area is subject to both the pulse duration and the electric-field envelope, Rabi oscillations of an ultra-short time scale can be implemented by ultrafast optical interaction at a strong-enough laser intensity regime. However, the spatial extent of the laser beam over the laser-atom interaction region inevitably causes spatial average effect that often leads to vanishing of the oscillatory behavior. To overcome this problem, homogenizing the spatial profile of laser beams~\cite{JLim2p, Weidemuller} and adapting chirped laser interaction~\cite{Milner} have been considered.

This paper aims quantitative analysis of spatially averaged Rabi oscillation. For this, we use the atom ensemble localized in a magneto-optical trap (MOT)~\cite{MOT} interacted with ultrafast laser pulses. As a theoretical model to investigate the spatially inhomogeneous interaction, we consider a Gaussian laser beam propagating along $z$ direction. The pulse area in Eq.~\eqref{SARO} is then represented in the polar coordinate system as 
\begin{eqnarray}
\Theta(r,z) =  \Theta_o\frac{w_o}{w(z)} e^{-{r^2}/{w(z)^2}} = \Theta_z e^{-{r^2}/{w(z)^2}},
\label{Theta_rz}
\end{eqnarray}
where $r=\sqrt{x^2+y^2}$, $w(z)$ is the beam waist at $z$, $w_o=w(0)$ is the minimal beam waist,
$\Theta_o$ is the maximal pulse area, and $\Theta_z=w_o \Theta_o /w(z)$. When we assume the atom density profile in the MOT is also a Gaussian, {\it i.e.,} 
$\rho(r,z) = \rho_o e^{-(r^2+z^2)/{w_a^2 }}$, 
the excited-state atom probability averaged over the entire atom ensemble, which we may call the ensemble-atom Rabi oscillation (EARO), is then given by
\begin{eqnarray}
\langle P (\Theta_o) \rangle &=& {\int  P(r,z; \Theta_o) \rho(r,z) dV}/{\int \rho(r,z) dV} \nonumber \\
     &=& \frac{2}{\sqrt{\pi} w_a^3} \int_{-\infty}^\infty dz  \int_0^\infty dr r \sin^2\frac{\Theta}{2} e^{-({r^2+z^2})/{w_a^2}} \nonumber \\
&=&  \frac{2}{\sqrt{\pi} w_a^3} \int_{-\infty}^\infty dz w^2 e^{-{z^2}/{w_a^2}} f(\Theta_z), 
\label{EARO}
\end{eqnarray}
where $f(\Theta_z)= \int_0^{\Theta_z} \left({\Theta}/{\Theta_z}\right)^{{w_a^2}/{\omega^2}} \sin^2 ({\Theta}/{2})  d \ln{\Theta}$. 

\begin{figure}[t]
\centerline{\includegraphics[width=8cm]{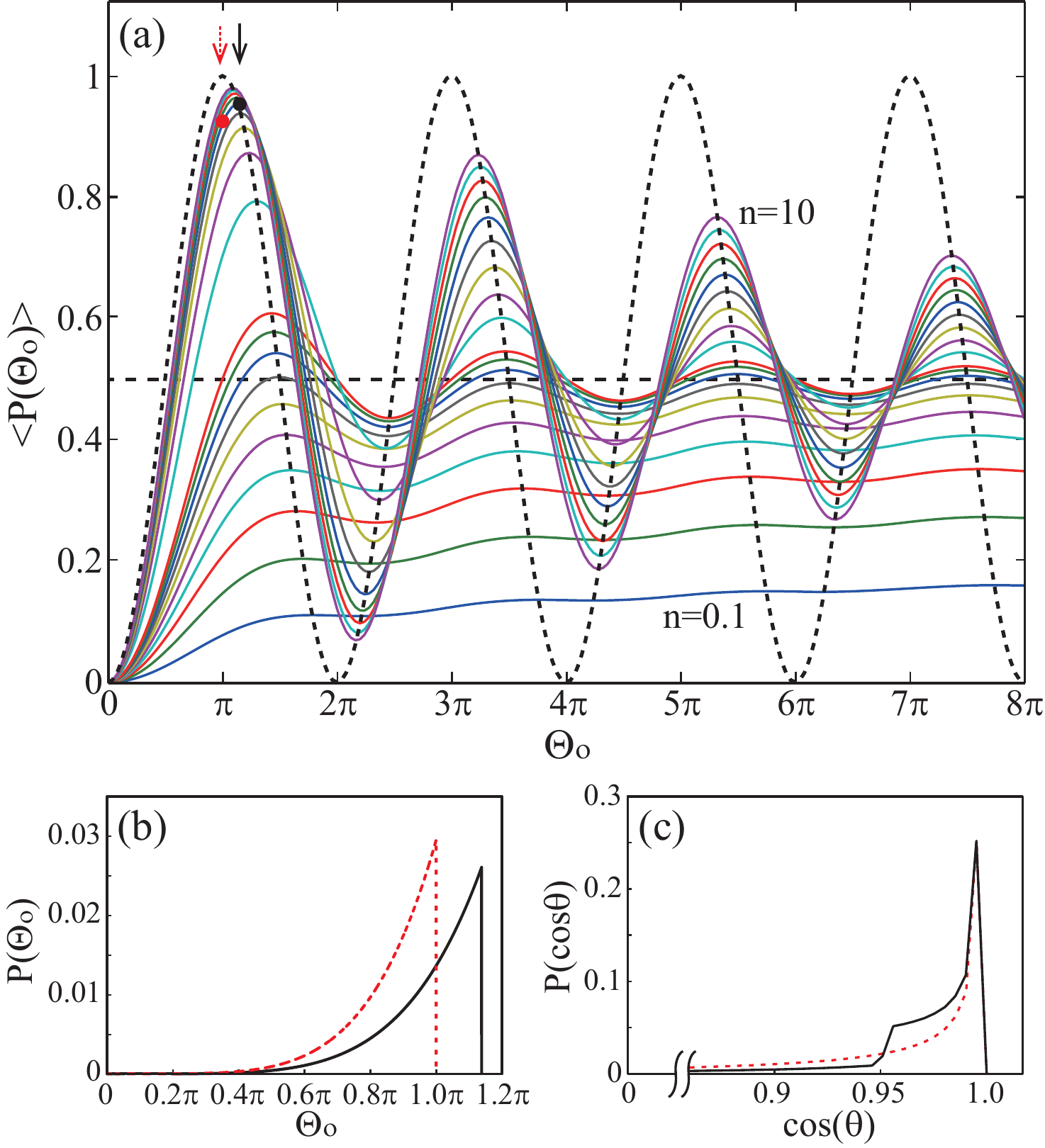}}
\caption{ (Color online) (a) Ensemble-atom Rabi oscillation in Eq.~\eqref{EARO} for various size ratios $w_o/w_a=\sqrt{n}$ for $n=0.1, 0.2, \cdots, 0.9$ and $1, 2. \cdots, 10$ (from the bottom to the top). Dotted line represents the single-atom Rabi oscillation in Eq.~\eqref{SARO}. (b,c) Atom probability distributions, at the marked points from the EARO curve for $w_o/w_a=\sqrt{6}$ in (a), plotted as a function of (b) $\Theta(r,z; \Theta_o)$, the pulse area, and (c) $\theta$, the polar angle of Bloch vector.
} \label{fig1}
\end{figure}

Figure~\ref{fig1} shows the numeriacl calculation of EARO in Eq.~\eqref{EARO} for various size ratios $w_o/w_a$, which is compared with SARO in Eq.~\eqref{SARO}. Note that the all EARO peaks coincide with the SARO curve in Fig.~\ref{fig1}. The locations of the EARO peaks are found from the condition 
\begin{eqnarray}
\frac{d \langle P\rangle}{d \Theta_o} = \frac{1}{\sqrt{\pi} w_a^3} \int_{-\infty}^\infty dz w^2 e^{-{z^2}/{w_a^2}} f' \frac{d \Theta_z}{d\Theta_o}=0.
\end{eqnarray}
and it is straightforward to show that $f(\Theta_n)=({w_a}/{w})^2 \sin^2{\Theta_n}/{2}$ for those $\Theta_n$ that satisfy $f'=0$. Therefore, we get $\Theta=\Theta_n$, Eq.~\eqref{EARO} becomes 
\begin{equation}
\langle P(\Theta_n) \rangle = \sin^2\frac{\Theta_n}{2},
\end{equation}
indicating that all EARO peaks are on the SARO curve.


\begin{figure}[b]
\centerline{\includegraphics[width=0.5\textwidth]{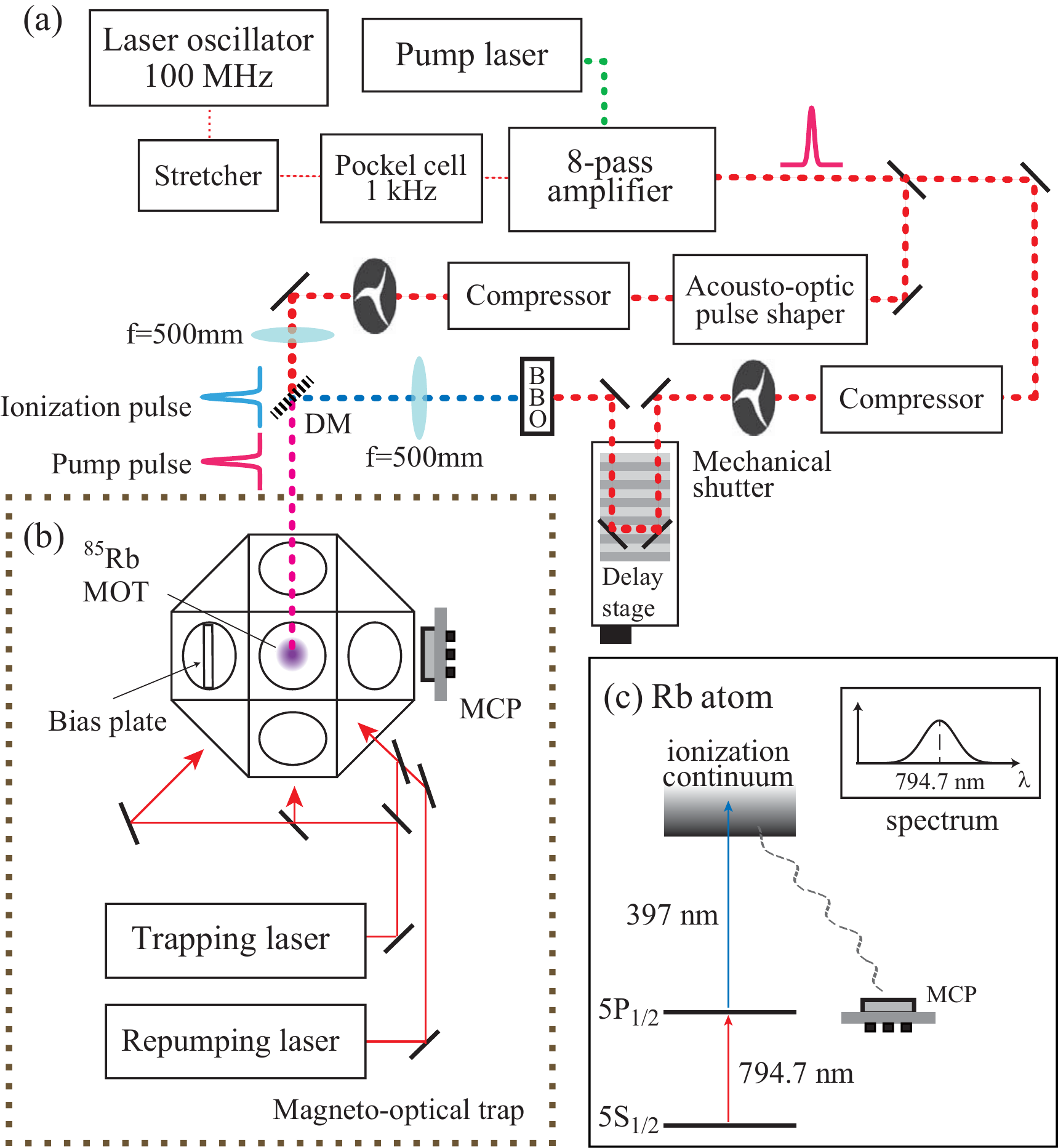}}
\caption{(a) Schematic diagram of the experimental setup. Ultrafast laser pulses were split into two pulses, one for Rabi oscillation and the other frequency-doubled for atom ionization. Both pulses were independently focused and delivered to the MOT by a dichroic mirror (DM). (b) Schematic diagram of the $^{85}$Rb MOT chamber. The trapping and re-pumping laser beams were adjusted to vary the atom cloud from 250 to 500~$\mu$m~\cite{MOTsize}. (c) Energy level diagram of the Rb atom and the laser spectrum. Atoms in the excited 5P$_{1/2}$ state were photo-ionized and  $Rb^+$ ions were transferred by bias electric plates and measured by a micro-channel plate detector (MCP).} \label{fig2}
\end{figure}

Experiments were performed with atomic rubidium ($^{85}$Rb) in a MOT~\cite{Sangkyung, JongseokSR2014} as shown in Figs.~\ref{fig2}(a,b). The  5S$_{1/2}$  and 5P$_{1/2}$ energy states are the ground and excited states, respectively, of the two-level system. The atoms were initially prepared in $F=3$ hyperfine level of 5S$_{1/2}$ by the MOT, and a $\pi$-polarized laser pulse induced $\Delta m_F=0$ transition to $F'=2$ and $3$ of 5P$_{1/2}$. The excited and ground states of the combined hyperfine levels formed an effective two-level system, for a ultrafast laser interaction with a moderated laser bandwidth~\cite{JongseokSR2014, Hyosub}.The atomic transition was driven by ultrafast laser pulses from a Ti:sapphire laser amplifier that produced 250-fs-short pulses at a repetition rate of 1~kHz. When the laser pulses were focused by a lens of focal length $f=500$~mm on to the atom ensemble, the pulse energy of up to 20~$\mu$J corresponded to the pulse area $\Theta_o$ up to $5\pi$. The laser spectrum was centered at $\lambda=794.7$~nm, the resonant wavelength of the 5S$_{1/2}$ $\to$ 5P$_{1/2}$ transition, and the spectral bandwidth was $\Delta \lambda=3$~nm (FWHM). The laser pulse was focused on the atom cloud by a $f = 500$~mm lens, and the beam size at the atom cloud was adjusted by translating the lens. The detection of the excited atom population was carried out by photo-ionization as shown in Fig.~\ref{fig2}(c). The probing UV pulse for the photo-ionization was prepared by frequency-doubling of a fraction of the main pulse via second-harmonic generation.The beam size of the probing UV pulse was adjusted by another lens to have twice the size of the main beam. Both laser pulses were combined after the lenses by a dichroic mirror and collinearly delivered to the atom cloud. The time difference controlled by a delay stage between the main and probing pulses was fixed to 10~ps, a thousand times smaller than the decay time of the $Rb$ excited state~\cite{Rbnumbers}.

\begin{widetext}

In each cycle of experiment operating at 2~Hz, atoms were first prepared by turning on the MOT for 500~ms by mechanical shutters, then immediately interacted with the ultrafast laser pulse, and finally photo-ionized by the probing UV pulse. The excited-state probability was estimated by comparing the fluorescence image counting of the atoms in the MOT and the ion count. 
Figure~\ref{fig3} shows the main experimental result, which clearly exhibits the seemingly decay-like oscillatory behavior. The above analysis on the ensemble-atom laser interaction predicts that such behavior is the spatially averaged Rabi oscillation. The agreement of the numerical calculation by Eq.~\ref{EARO} and the experimental result is excellent. It is noted, however, that the discrepancy between them is evident in particular for a high pulse-area exceeding $\Theta_o=3\pi$ and also for a higher spatial inhomogeneity in Fig.~\ref{fig3}(c) for $w_o/w_a=1$ than the others. As a cause of the error, we can consider the three-photon ionization directly by the main laser pulse in addition to the one-photo ionization by the probing UV pulse. Such effect is however already systematically taken into account in the data analysis, and the error is estimated less than 2\% in the given range of pulse area. The main reason for the discrepancy is the axis mis-alignment of at most 50$\mu$m between the laser and the atom cloud, not to mention the imperfect shape of the atom cloud. Our calculation predicts the case in Fig.~3(c) exhibits total 5\% of error. Furthermore, the ratios of the adjacent peaks are uniquely determined as a function of the size ratio $w_o/w_a$, which result can be used as an alternative means to calibrate the excited-state probability of the atom ensemble.
\begin{figure}[h]
    \centerline{\includegraphics[width=\textwidth]{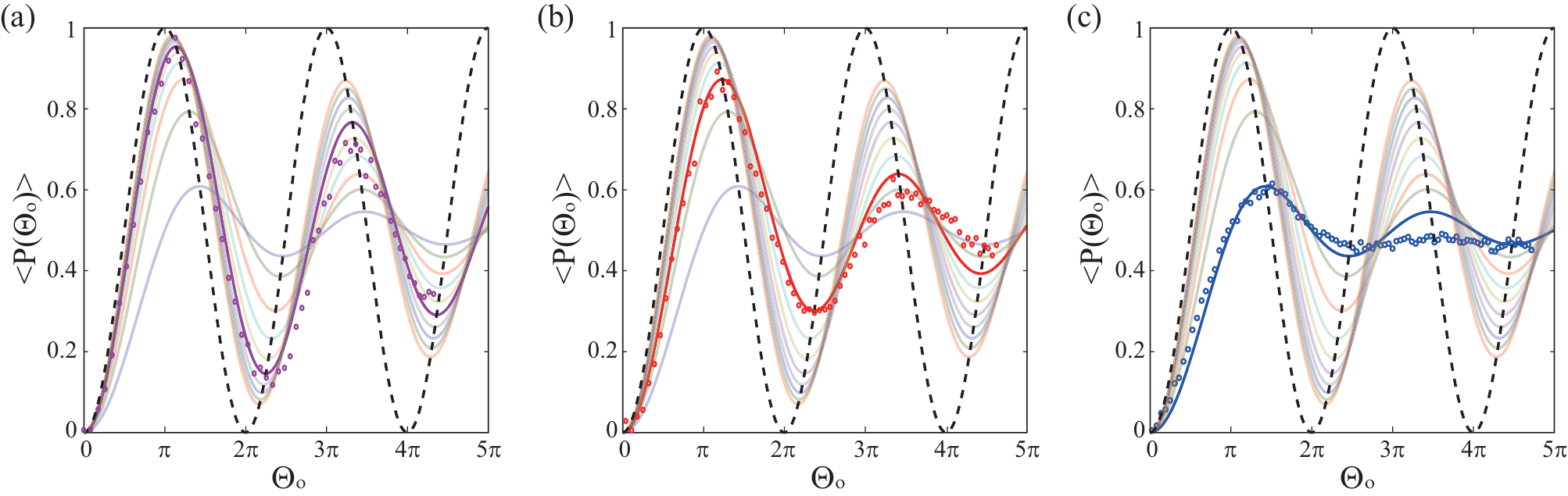}}
    \caption{Experimental result of ensemble-atom Rabi oscillations: (a) Laser beam width ($w_o$) was 2.5 times of the atom cloud size ($w_a$) or $w_o= 2.5 w_a$, (b) $w_o= 1.7 w_a$, and (c) $w_o = w_a$. The highlighted line in each figure illustrates the calculation for the closest integer $(w_o/w_a)^2$ that corresponds to (a) $(w_o/w_a)^2=6$ , (b) ${3}$ , and (c)  1, respectively.} 
    \label{fig3}
\end{figure}

In the second experiment, we considered a composite pulse to achieve a higher-fidelity Rabi oscillation.  We tested the three-pulse composite consisting of two $\pi/2$ rotations about $x$-axis and a $\pi$ rotation about $y$-axis, or  $R_x(\pi/2)R_y(\pi)R_x(\pi/2)$, which sequence of pulses is well-known in NMR designed to correct errors caused by pulse-area fluctuation~\cite{CompositePulse, RakreungdetPRA2009}.
The $\pi$-rotation $R_y(\pi)$ about the $y$-axis in the middle corrects the rotation error of the pair of $\pi/2$ rotations $R_x(\pi/2)^2$. In our experiment, we used the three-pulse composite to reduce the spatial inhomogeneity in the ensemble-atom laser interaction. To make the three pulses of the specific amplitude and phase coding, we used an acousto-optic pulse shaper as shown in Fig.~\ref{fig2}(a), and the relative amplitudes of the pulses were checked by in-situ auto-correlation measurement~\cite{FROG}. 
Figure~\ref{fig4} shows the result of the pulse composite experiment. The first-order corrections of the pulse area $\Theta_o$ for the $\pi$ and $\pi/2$ pulses in an ensemble-atom experiment are respectively given by $\pi+\alpha$ and $\pi/2+\beta$, where $\alpha$ and $\beta$ are determined by the size ratio $w_o/w_a$. The experiment was thus performed by a pulse sequence $R_x(\pi/2+\alpha)R_y(\pi+\beta)R_x(\pi/2+\alpha)$, and the excited-state population is plotted in Fig.~\ref{fig4} as a function of $\Theta_o=\pi+2\alpha=\pi+\beta$ by fixing $\beta=2\alpha$ for the experimental convenience. The result in Fig.~\ref{fig4} clearly demonstrates 15\% of increase at the first peak of the oscillation by the composite pulse  (black circles) compared to the oscillation by the single pulse (red boxes). The robustness of the composite pulse scheme manifested by the broadened oscillation peak around the $\pi$ rotation is also clearly observed in Fig.~\ref{fig4}. It is straight-forward to show that the excited state probability for a single-atom excitation is given by 
\begin{equation}
P_e(\Theta_o)=1-\cos^4\frac{\Theta_o}{2} \end{equation} 
for the pulse composite pulse, which results in a broader peak shape around $\Theta_o=\pi$ than $\sin^2{(\Theta_o/2)}$ for a single pulse in Eq.~\ref{SARO}. 
\end{widetext}

\begin{figure}[t]
    \centerline{\includegraphics[width=8cm]{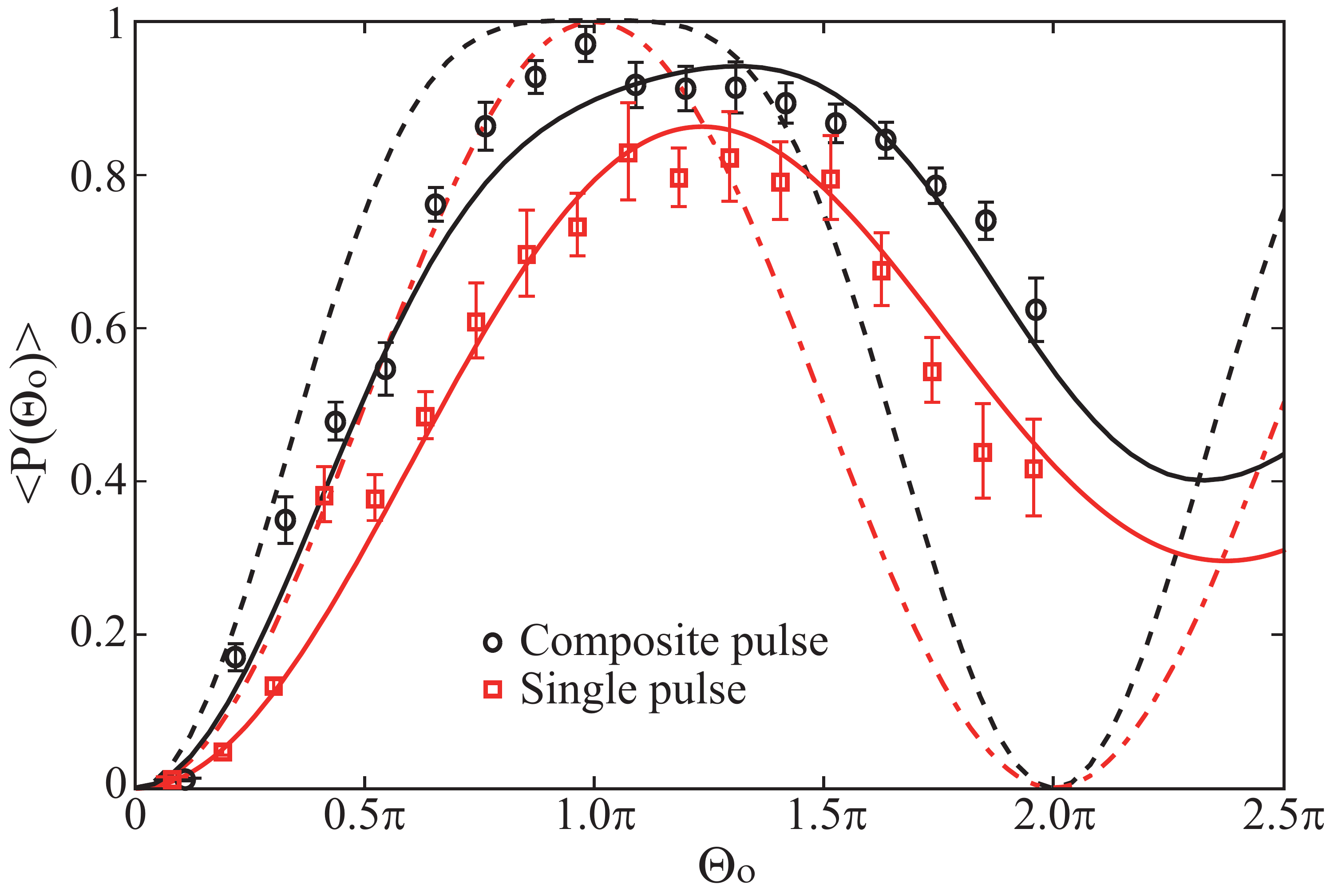}}
    \caption{(Color online) Composite-pulse experiment of ensemble-atom Rabi oscillation. For an atom ensemble of the size 1.7 times smaller than a laser beam ({\it i.e.}, $w_o=1.7 w_a$), the excited-state population for the composite-pulse operation $R_x(\Theta_o/2)R_y(\Theta_o)R_x(\Theta_o/2)$ was measured and plotted in black circles. In comparison, the single pulse experiment $R_x(\Theta_o)$ was plotted in red boxes. The solid lines represent the corresponding numerical calculations, when the spatial inhomogeneity of the ensemble-atom experiment is taken into account. The dotted lines are for the spatially homogeneous case ({\it i.e.}, $w_o \gg  w_a$), when the theoretical formulas are given by $1-\cos^4{(\Theta_o/2)}$ (black) for the composite pulse and $\sin^2{(\Theta_o/2)}$ (red) for single pulses, respectively.} \label{fig4}
\end{figure}

In summary, we have studied Rabi oscillation of a spatially-confined atom ensemble with a Gaussian laser beam. Based on theoretical analysis, we have found that that the peak positions of the ensemble-atom Rabi oscillation are uniquely determined by the size ratio between the atom ensemble and the laser beam, and the result has been confirmed by the ultrafast laser experiment with cold atom clouds. The reduced fidelity of the atom-ensemble Rabi-flopping has been compensated by the proof-of-principle demonstration of the three-pulse composite operation $R_x(\pi/2)R_y(\pi)R_x(\pi/2)$.  

This research was supported in part by Basic Science Research Program [2013R1A2A2A05005187] through the National Research Foundation of Korea and in part by Samsung Science and Technology Foundation [SSTF-BA1301-12].

\end{document}